\documentclass[twocolumn,aps,pra,floatfix,showpacs,noshowkeys,epsfig,graphics]{revtex4}%
\usepackage{graphicx}
\usepackage{amsmath}
\usepackage{amsfonts}
\usepackage{amssymb}
\usepackage{amsthm}
\newtheorem{theorem}{Theorem}

\newtheorem{definition}{Definition}
\newcommand{\newc}{\newcommand}

\newcommand{\bea}{\begin{eqnarray}}
\newcommand{\eea}{\end{eqnarray}}
\newcommand{\ket}[1]{| #1 \rangle}
\newcommand{\bra}[1]{\langle #1 |}

\newcommand{\tr}{\textrm{tr}}

\newc{\beq}    {\begin{equation}}
\newc{\eeq}    {\end{equation}}
\newc{\beqa}    {\begin{eqnarray}}
\newc{\eeqa}    {\end{eqnarray}}
\newc{\bs}    {\section}
\newc{\no}    {\\ \nonumber}

\begin{document}
\title{Deciding whether a quantum state has secret correlations is an NP-complete problem }
\author{Jae-Weon Lee}\email{scikid@kias.re.kr}
\author{DoYong Kwon}\email{doyong@kias.re.kr}
\author{Jaewan Kim}\email{jaewan@kias.re.kr}
\affiliation{School of Computational Sciences,
             Korea Institute for Advanced Study, Seoul 130-722, Korea}
\date{\today}
\begin{abstract}
From the NP-hardness of the quantum separability problem and the relation between
 bipartite entanglement and the secret key correlations,
it is shown that the problem deciding whether a given  quantum
state has secret correlations in it or not is in NP-complete.
\end{abstract}
\maketitle

Recent progress in theories and
 experiments on quantum key distribution
(QKD)
allows one to think QKD as the first successful application of  quantum information science\cite{gisin}.
However, minimal and essential physical ingredients for QKD are still not clear.
For example, the equivalence between bipartite entanglement and secret key generation is still unproved\cite{problem}.
Acin et al. \cite{acin:167901} showed that,  under the assumption that legitimate parties measure
 only single copies of the state and
eavesdropper performs individual attack, secret bits can be asymptotically distilled
from any two-qubit entangled state.
Recently, in Ref. \cite{acin:020501,curty:217903} it was shown that entangled states can be mapped
 into classical probability distributions containing secret correlations
and vice versa.
It was also shown that, surprisingly, even from bound entangled states one can distill an arbitrarily secure key
\cite{horodecki:160502}.
All these results give rise to a fundamental question about
 exact connections between
 entanglement of a quantum state and the private
key distillable from the  state \cite{augusiak:010305}.
On the other hand, in computational science there is a long standing open problem
called $P$ vs. $NP$ problem; {\it{Is an easy checkable problem always easy solvable}}\cite{Papadimitriou}?
Many practical classical cryptography systems such as RSA\cite{rsa} and Elliptic curve cryptography\cite{ecc1,ecc2} rely on
 difficulty of some mathematical problems in $NP$ class for security, while security for quantum key distribution
(QKD) systems relies on physical laws.
In this paper a relation between this famous computational complexity problem and
quantum key distribution is investigated. More precisely,
we show that  deciding whether a given quantum state
 has secret correlations  ({\em i.e.}, $I_{form}(X;Y|Z)>0$, see below)  in it or not is in NP-complete class.

Let us begin by shortly reviewing the secret key generation from a given $classical$
distribution $P(X,Y,Z)$ of random variables $X,Y$, and $Z$.
This distribution might have been obtained from measurements of  shared states independently
done by legitimate parties, Alice($M_X$) and Bob($M_Y$), and an eavesdropper Eve($M_Z$).
Then, for a given $P(X,Y,Z)$
  a secret key rate $S(X;Y||Z)$ is the maximum key generation rate
   from the distribution by local operations and
public classical communication(LOPC). Similarly, one can define
  the information of formation $I_{form}(X;Y|Z)$ which is the amount of secret bits  needed for preparing
$P(X,Y,Z)$.
  They satisfy a relation\cite{maurer}
\beq
S(X;Y||Z)\leq I(X;Y|Z)_{form},
\eeq
which states that as the entanglement
cost is  larger than or equal to distillable entanglement,
 so the amount of secret bits  needed for preparing the distribution
is  larger than or equal to the amount of secret bits that is distillable  from it.
There is a well known following theorem on the relation between bipartite entanglement of a state
and the secret correlations in it.
\begin{theorem}[Equivalence of bipartite entanglement and secret correlation]\cite{acin:020501,curty:217903}
\label{acingisin}
  Let $|\psi_{ABE}\rangle$ be a pure quantum state shared by Alice, Bob, and
Eve, such that the state is a purification of Alice and Bob's bipartite density matrix
$\rho_{AB}$ ({\em i.e.}, $\rho_{AB}=tr_E(\ket{\psi_{ABE}}\bra{\psi_{ABE}})$. Then, $\rho_{AB}$ is entangled if and only if there exist
measurements of $|\psi_{ABE}\rangle$ by Alice ($M_X$) and Bob ($M_Y$), such that for
any measurement by Eve ($M_Z$), the corresponding probability
distribution $P(X,Y,Z)$ contains secret correlations,{\em i.e.},
$I_{form}(X; Y|Z)>0$.
\end{theorem}
This theorem is proven by showing the existence of an entanglement witness from the measurement
operators $M_X$ and $M_Y$.
In this paper we will consider only bipartite states in $\mathcal{H}_A \otimes \mathcal{H}_B $
where $dim(\mathcal{H}_A)=M$ and $dim(\mathcal{H}_B)=N$.

Since Turing machines can not represent arbitrary real or complex numbers
from now on we  deal with only density matrices  of which representations $[\rho]$ have rational entries with finite precision.
Now we define a problem  deciding whether a given state has secret correlations.
\begin{definition}[Quantum Secret Correlation problem(QSCORR)]
\label{Secretcorrelation}
Let $[\rho_{AB}]$ be a rational bipartite mixed state
having a  purification
 $\ket{\psi_{ABE}}$ as described in  Theorem \ref{acingisin}.
Given $[\rho_{AB}]$, does any $P(X,Y,Z)$ from $\ket{\psi_{ABE}}$
contain secret correlations, that is, $I_{form}(X;Y|Z)>0$?
\end{definition}

To tackle this problem we need the famous theorem by Gurvits\cite{gurvits-2003,1039332} about deciding
entanglement of a given density matrix
on a (deterministic) Turing machine ({\em i.e.}, an abstraction of ordinary computers).
To understand the theorem let us recall some definitions in computational complexity theory\cite{Papadimitriou}.
We say
that a problem $A$
is polynomially reducible to another problem $B$
if there exists a polynomial-time algorithm that
converts each input(instance) $I_A$ of $A$ to
 another input $I_B$ of $B$ such that
$I_A$ is a $yes$-instance of $A$
if and only if $I_B$ is a $yes$-instance of $B$.
In this case we denote this relation as $A\leq_P B$.
The NP (Non-deterministic Polynomial time) class is the set of decision problems
 that can be verified by a
  Turing machine in polynomial time. Many practical and important problems
  such as the factoring (a decision version)  and the graph isomorphism problem belong to this class.
  The NP-hard class is the class of all problems $B$
    such that for every problem
     in NP there exists a polynomial time reduction to  $B$.
     Many interesting physical problems belong to this class\cite{eisert-2006}.
 The NP-complete class is an intersection of the NP class and the NP-hard class.

 One can naturally imagine the following separability problem of rational density matrices.
\begin{definition}[Rational quantum separability problem (EXACT QSEP)]
 Given a bipartite rational $[\rho]$, is $[\rho]$
separable?
\end{definition}

Unfortunately, EXACT QSEP  encounters a mathematical difficulty near the boundary of $S_{M,N}$\cite{ioannou-2006,guhne:170502}
about representing  density matrices with rational numbers.
Here $S_{M,N}$ is a convex set of separable density matrices acting on $\mathcal{H}_A \otimes \mathcal{H}_B $.
Instead, Gurvits
considered a problem  asking whether a given $[\rho]$ is close to separable states\cite{gurvits-2003}.
\begin{definition}[Weak membership problem (WMEM)]
Given a rational vector $[\rho]$ and a rational
$\delta>0$, assert that  either
\begin{eqnarray}
\left[\rho\right] &\in& S(S_{M,N},\delta), ~\text{or}\\
\left[\rho\right] &\notin& S(S_{M,N},-\delta),
\end{eqnarray}
where  $S(S_{M,N},\delta)$ is a union of all $\delta$-balls of which centers belong to
$S_{M,N}$ and $S(S_{M,N},-\delta)$ is
a set of centers of $\delta$-balls where the $\delta$-balls are contained  in $S_{M,N}$.
\end{definition}

Deciding, quantifying and distillating entanglement are subjects of intensive investigations in quantum
information community\cite{bruss-2002-43}.
For example, an improved algorithm for quantum separability and entanglement detection on classical computers
is suggested\cite{ioannou:060303} and
Doherty {\em et al.} constructed families of operational criteria for separability
 based on semidefinite programs\cite{PhysRevLett.88.187904}.
 Despite all these efforts an efficient ({\em i.e.}, polynomial time) algorithm for the separability problem is still unknown.
The  following seminal theorem due to Gurvits explains why the quantum separability problem is so hard.

\begin{theorem}[Gurvits]
\label{gurvits}
 WMEM for $S_{M,N}$ is NP-hard
with respect to the complexity-measure $(N +
<[\rho]> + <\delta>)$  if $N\leq M\leq \frac{N(N-1)}{2}+2$, where
$<>$ denotes the size of the encoding.
\end{theorem}
He demonstrated a polynomial time reduction from an
 NP-complete problem called  KNAPSACK to WMEM($S_{M,N}$) after a series of transformations.

At first glance, it might seem that
knowing Theorem \ref{acingisin} and Theorem \ref{gurvits}
one can easily prove the NP-completeness of the problem deciding whether a given
state has secret correlations(QSCORR).
But real situation is complicated.
To be proved as an NP-complete problem, the problem should be a decision problem.
However,
WMEM is not a decision problem,
because inputs corresponding to states near the
boundary of $S_{M,N}$ can give both possible answers\cite{ioannou-2006}.
To avoid this ambiguity  Ioannou designed a decidable separability problem called
QSEP\cite{ioannou-2006} asking whether, given a rational density operator $[\rho]$,
there exists a separable density operator $\sigma$ close to $[\rho]$.
\begin{definition}[QSEP] \label{qsep} Given a rational bipartite density matrix
 $[\rho]$ acting on $\mathcal{H}_A \otimes \mathcal{H}_B $,
 and $p$-bit  rational numbers
$\epsilon$ and $\delta'$; does there exist a separable state
$\sigma=\sum_{i=1}^{M^2N^2} p_i
\ket{\alpha_i}\bra{\alpha_i}\otimes
\ket{\beta_i}\bra{\beta_i}$,
of which $p$-bit truncated and unnormalized version
$\tilde{\sigma}=\sum_{i=1}^{M^2N^2} \tilde{p}_i
\ket{\tilde{\alpha}_i}\bra{\tilde{\alpha}_i}\otimes
\ket{\tilde{\beta}_i}\bra{\tilde{\beta}_i}$
satisfying \\
$i)~ |[\rho] -\sigma|_2 < \delta'$, and \\
$ii)~ |\sigma-\tilde{\sigma}|_2<\epsilon$?\\
Here $|A-B|_2\equiv\sqrt{\tr\left((A -B)^2\right)}$, $\tilde{p}_i\ge 0$ is a $p$-bit rational number
and $p_i\ge 0$.
\end{definition}
We have adopted a slightly modified definition from the original one of
QSEP in \cite{ioannou-2006} for our purpose, but basically
two definitions are equivalent.
\begin{theorem}
\label{thqsep}
QSEP is in NP-complete\cite{ioannou-2006}.
\end{theorem}
The NP-completeness of QSEP was proven by reduction from
WMEM. QSEP is carefully designed so that
for an instance $I([\rho],\delta)$ of WMEM one call QSEP with an instance
$I'([\rho],p,\epsilon,\delta')$ such that $\delta\ge \delta'+\epsilon$.
To utilize this definition we
consider a negation of QSCORR with  error.
\begin{definition}[No Quantum Secret Correlation (NQSCORR)]
\label{Secretcorrelation}
Given a rational bipartite density matrix
 $[\rho]$,
 does there exist a  state
$\sigma$,
satisfying \\
$i)~ |[\rho] -\sigma|_2 < \delta'$, \\
$ii)~ |\sigma-\tilde{\sigma}|_2<\epsilon$, and \\
$iii)$ for any  purification
 $\ket{\psi_{ABE}}$ of $\sigma$ as described in Theorem \ref{acingisin},
it contains  no secret correlations, that is, $I_{form}(X;Y|Z)=0$?
(Here $\delta'$,$\epsilon$ are $p$-bit rational  numbers and
$\tilde{\sigma}$ is a $p$-bit truncation of $\sigma$)
\end{definition}
Note that in the zero-error limit ($\delta'\rightarrow 0,\epsilon\rightarrow 0 )$
this problem reduces to the exact negation of $QSCORR$.
\begin{theorem}
\label{qscorr}
NQSCORR is in NP-complete.
\end{theorem}
\begin{proof}\label{SecretcorrelationNPcompleteProof}
Basically, this theorem is a corollary of Theorem \ref{acingisin} and Theorem \ref{thqsep}.
If there is an algorithm that solves NQSCORR, then one can call the
algorithm to solve QSEP.
More precisely, given $I([\rho],p,\delta',\epsilon)$ of QSEP one can call  NQSCORR with $I'([\rho],p,\delta',\epsilon)$.
NQSCORR returns $yes$ if and only if QSEP returns $yes$  because of the equivalence
of bipartite entanglement and secret correlations (Theorem \ref{acingisin}).
This means NQSCORR is at least as hard as QSEP which is in NP-complete class.
Therefore NQSCORR is also in NP-hard. Furthermore,
given a certificate ($\tilde{\sigma},M_X,M_Y,M_Z)$ one can quickly ({\em i.e.}, in a polynomial time) verify whether $I_{form}(X,Y|Z)$
from $P(X,Y,Z)$ is positive or not. Hence NQSCORR is also in NP.
Therefore, NQSCORR is in NP-complete class.
\end{proof}
The full reduction chain is
$KNAPSACK\le_P RSDF \le_P WVAL \le_P WMEM \le_P QSEP \le_P NQSCORR$
(See \cite{gurvits-2003,ioannou-2006} for definitions of the intermediate problems).

One may think of another related  and more interesting problem asking
whether a given bipartite density matrix has non-zero secret key generation rate,
that is, $S(X,Y||Z)>0$.
Since $I_{form}(X;Y|Z)>0$ is not a sufficient condition but a
 necessary condition for $S(X,Y||Z)>0$, ({\em i.e.}, there is a bound information\cite{acin:107903}),
 we could not answer to this interesting question within our approach.

 What our results imply is that
 there is no easy procedure or simple formula for deciding whether
 a given quantum state gives rise to secret correlations if $P \neq NP$ (which is usually believed).
Conversely, as a byproduct of our results, if one can find a polynomial time algorithm solving the NQSCORR
 problem on a deterministic Turing machine
it means $P=NP$.
Our results also reveal that the P vs. NP problem is not only related to
 classical cryptography  but also to quantum cryptography
 in a different way.

\section*{Acknowledgments}
We thank Joonwoo Bae and  Sung-il Pae
for helpful discussions.
J. Lee and
J. Kim
was supported by the Korea Ministry of Information and Communication
with the``Next Generation Security Project".



\begin{thebibliography}{50}
\expandafter\ifx\csname natexlab\endcsname\relax\def\natexlab#1{#1}\fi
\expandafter\ifx\csname bibnamefont\endcsname\relax
  \def\bibnamefont#1{#1}\fi
\expandafter\ifx\csname bibfnamefont\endcsname\relax
  \def\bibfnamefont#1{#1}\fi
\expandafter\ifx\csname citenamefont\endcsname\relax
  \def\citenamefont#1{#1}\fi
\expandafter\ifx\csname url\endcsname\relax
  \def\url#1{\texttt{#1}}\fi
\expandafter\ifx\csname urlprefix\endcsname\relax\def\urlprefix{URL }\fi
\providecommand{\bibinfo}[2]{#2}
\providecommand{\eprint}[2][]{\url{#2}}

\bibitem[{\citenamefont{Gisin et~al.}(2002)\citenamefont{Gisin, Ribordy,
  Tittel, and Zbinden}}]{gisin}
\bibinfo{author}{\bibfnamefont{N.}~\bibnamefont{Gisin}},
  \bibinfo{author}{\bibfnamefont{G.}~\bibnamefont{Ribordy}},
  \bibinfo{author}{\bibfnamefont{W.}~\bibnamefont{Tittel}}, \bibnamefont{and}
  \bibinfo{author}{\bibfnamefont{H.}~\bibnamefont{Zbinden}},
  \bibinfo{journal}{Rev. Mod. Phys.} \textbf{\bibinfo{volume}{74}},
  \bibinfo{pages}{145} (\bibinfo{year}{2002}).

\bibitem[{\citenamefont{Horodecki et~al.}(2006)\citenamefont{Horodecki,
  Horodecki, Horodecki, and Oppenheim}}]{problem}
\bibinfo{author}{\bibfnamefont{K.}~\bibnamefont{Horodecki}},
  \bibinfo{author}{\bibfnamefont{M.}~\bibnamefont{Horodecki}},
  \bibinfo{author}{\bibfnamefont{P.}~\bibnamefont{Horodecki}},
  \bibnamefont{and}
  \bibinfo{author}{\bibfnamefont{J.}~\bibnamefont{Oppenheim}},
  \bibinfo{journal}{quant-ph/0506189}  (\bibinfo{year}{2006}).

\bibitem[{\citenamefont{Acin et~al.}(2003)\citenamefont{Acin, Masanes, and
  Gisin}}]{acin:167901}
\bibinfo{author}{\bibfnamefont{A.}~\bibnamefont{Acin}},
  \bibinfo{author}{\bibfnamefont{L.}~\bibnamefont{Masanes}}, \bibnamefont{and}
  \bibinfo{author}{\bibfnamefont{N.}~\bibnamefont{Gisin}},
  \bibinfo{journal}{Phys. Rev. Lett.} \textbf{\bibinfo{volume}{91}},
  \bibinfo{eid}{167901} (\bibinfo{year}{2003}).

\bibitem[{\citenamefont{Acin and Gisin}(2005)}]{acin:020501}
\bibinfo{author}{\bibfnamefont{A.}~\bibnamefont{Acin}} \bibnamefont{and}
  \bibinfo{author}{\bibfnamefont{N.}~\bibnamefont{Gisin}},
  \bibinfo{journal}{Phys. Rev. Lett.} \textbf{\bibinfo{volume}{94}},
  \bibinfo{eid}{020501} (\bibinfo{year}{2005}).

\bibitem[{\citenamefont{Curty et~al.}(2004)\citenamefont{Curty, Lewenstein, and
  L\"{u}tkenhaus}}]{curty:217903}
\bibinfo{author}{\bibfnamefont{M.}~\bibnamefont{Curty}},
  \bibinfo{author}{\bibfnamefont{M.}~\bibnamefont{Lewenstein}},
  \bibnamefont{and}
  \bibinfo{author}{\bibfnamefont{N.}~\bibnamefont{L\"{u}tkenhaus}},
  \bibinfo{journal}{Phys. Rev. Lett.} \textbf{\bibinfo{volume}{92}},
  \bibinfo{eid}{217903} (\bibinfo{year}{2004}).

\bibitem[{\citenamefont{Horodecki et~al.}(2005)\citenamefont{Horodecki,
  Horodecki, Horodecki, and Oppenheim}}]{horodecki:160502}
\bibinfo{author}{\bibfnamefont{K.}~\bibnamefont{Horodecki}},
  \bibinfo{author}{\bibfnamefont{M.}~\bibnamefont{Horodecki}},
  \bibinfo{author}{\bibfnamefont{P.}~\bibnamefont{Horodecki}},
  \bibnamefont{and}
  \bibinfo{author}{\bibfnamefont{J.}~\bibnamefont{Oppenheim}},
  \bibinfo{journal}{Phys. Rev. Lett.} \textbf{\bibinfo{volume}{94}},
  \bibinfo{eid}{160502} (\bibinfo{year}{2005}).

\bibitem[{\citenamefont{Augusiak and Horodecki}(2006)}]{augusiak:010305}
\bibinfo{author}{\bibfnamefont{R.}~\bibnamefont{Augusiak}} \bibnamefont{and}
  \bibinfo{author}{\bibfnamefont{P.}~\bibnamefont{Horodecki}},
  \bibinfo{journal}{Phys. Rev. A} \textbf{\bibinfo{volume}{74}},
  \bibinfo{eid}{010305} (\bibinfo{year}{2006}).

\bibitem[{\citenamefont{Papadimitriou}(1994)}]{Papadimitriou}
\bibinfo{author}{\bibfnamefont{C.}~\bibnamefont{Papadimitriou}},
  \emph{\bibinfo{title}{Computational Complexity}}
  (\bibinfo{publisher}{Addison-Wesley, Newyork}, \bibinfo{year}{1994}).

\bibitem[{\citenamefont{Rivest et~al.}(1978)\citenamefont{Rivest, Shamir, and
  Adleman}}]{rsa}
\bibinfo{author}{\bibfnamefont{R.}~\bibnamefont{Rivest}},
  \bibinfo{author}{\bibfnamefont{A.}~\bibnamefont{Shamir}}, \bibnamefont{and}
  \bibinfo{author}{\bibfnamefont{L.}~\bibnamefont{Adleman}},
  \bibinfo{journal}{Communications of the ACM} \textbf{\bibinfo{volume}{21}},
  \bibinfo{pages}{120} (\bibinfo{year}{1978}).

\bibitem[{\citenamefont{Koblitz}(1987)}]{ecc1}
\bibinfo{author}{\bibfnamefont{N.}~\bibnamefont{Koblitz}},
  \bibinfo{journal}{Mathematics of Computation} \textbf{\bibinfo{volume}{48}},
  \bibinfo{pages}{203} (\bibinfo{year}{1987}).

\bibitem[{\citenamefont{Miller}(1985)}]{ecc2}
\bibinfo{author}{\bibfnamefont{V.}~\bibnamefont{Miller}},
  \bibinfo{journal}{CRYPTO 85}  (\bibinfo{year}{1985}).

\bibitem[{\citenamefont{Maurer and Wolf}(1999)}]{maurer}
\bibinfo{author}{\bibfnamefont{U.}~\bibnamefont{Maurer}} \bibnamefont{and}
  \bibinfo{author}{\bibfnamefont{S.}~\bibnamefont{Wolf}},
  \bibinfo{journal}{IEEE Trans. Inf. Theory} \textbf{\bibinfo{volume}{45}},
  \bibinfo{pages}{499} (\bibinfo{year}{1999}).

\bibitem[{\citenamefont{Gurvits}(2003)}]{gurvits-2003}
\bibinfo{author}{\bibfnamefont{L.}~\bibnamefont{Gurvits}},
  \bibinfo{journal}{Proceedings of the 35th ACM Symposium on Theory of
  Computing ACM Press, New York, 2003} p.~\bibinfo{pages}{10}
  (\bibinfo{year}{2003}), \eprint{quant-ph/0303055}.

\bibitem[{\citenamefont{Gurvits}(2004)}]{1039332}
\bibinfo{author}{\bibfnamefont{L.}~\bibnamefont{Gurvits}}, \bibinfo{journal}{J.
  Comput. Syst. Sci.} \textbf{\bibinfo{volume}{69}}, \bibinfo{pages}{448}
  (\bibinfo{year}{2004}).

\bibitem[{\citenamefont{Eisert}(2006)}]{eisert-2006}
\bibinfo{author}{\bibfnamefont{J.} \bibnamefont{Eisert}},
  \bibinfo{journal}{quant-ph/0609051}  (\bibinfo{year}{2006}).

\bibitem[{\citenamefont{Ioannou}(2006)}]{ioannou-2006}
\bibinfo{author}{\bibfnamefont{L.~M.} \bibnamefont{Ioannou}},
  \bibinfo{journal}{quant-ph/0603199}  (\bibinfo{year}{2006}).

\bibitem[{\citenamefont{Guhne and L\"{u}tkenhaus}(2006)}]{guhne:170502}
\bibinfo{author}{\bibfnamefont{O.}~\bibnamefont{Guhne}} \bibnamefont{and}
  \bibinfo{author}{\bibfnamefont{N.}~\bibnamefont{L\"{u}tkenhaus}},
  \bibinfo{journal}{Phys. Rev. Lett.} \textbf{\bibinfo{volume}{96}},
  \bibinfo{eid}{170502} (\bibinfo{year}{2006}).

\bibitem[{\citenamefont{Bruss}(2002)}]{bruss-2002-43}
\bibinfo{author}{\bibfnamefont{D.}~\bibnamefont{Bruss}},
  \bibinfo{journal}{Journ. Math. Phys.} \textbf{\bibinfo{volume}{43}},
  \bibinfo{pages}{4237} (\bibinfo{year}{2002}).

\bibitem[{\citenamefont{Ioannou et~al.}(2004)\citenamefont{Ioannou,
  Travaglione, Cheung, and Ekert}}]{ioannou:060303}
\bibinfo{author}{\bibfnamefont{L.~M.} \bibnamefont{Ioannou}},
  \bibinfo{author}{\bibfnamefont{B.~C.} \bibnamefont{Travaglione}},
  \bibinfo{author}{\bibfnamefont{D.}~\bibnamefont{Cheung}}, \bibnamefont{and}
  \bibinfo{author}{\bibfnamefont{A.~K.} \bibnamefont{Ekert}},
  \bibinfo{journal}{Phys. Rev. A} \textbf{\bibinfo{volume}{70}},
  \bibinfo{eid}{060303} (\bibinfo{year}{2004}).

\bibitem[{\citenamefont{Doherty et~al.}(2002)\citenamefont{Doherty, Parrilo,
  and Spedalieri}}]{PhysRevLett.88.187904}
\bibinfo{author}{\bibfnamefont{A.~C.} \bibnamefont{Doherty}},
  \bibinfo{author}{\bibfnamefont{P.~A.} \bibnamefont{Parrilo}},
  \bibnamefont{and} \bibinfo{author}{\bibfnamefont{F.~M.}
  \bibnamefont{Spedalieri}}, \bibinfo{journal}{Phys. Rev. Lett.}
  \textbf{\bibinfo{volume}{88}}, \bibinfo{pages}{187904}
  (\bibinfo{year}{2002}).

\bibitem[{\citenamefont{Acin et~al.}(2004)\citenamefont{Acin, Cirac, and
  Masanes}}]{acin:107903}
\bibinfo{author}{\bibfnamefont{A.}~\bibnamefont{Acin}},
  \bibinfo{author}{\bibfnamefont{J.~I.} \bibnamefont{Cirac}}, \bibnamefont{and}
  \bibinfo{author}{\bibfnamefont{L.}~\bibnamefont{Masanes}},
  \bibinfo{journal}{Phys. Rev. Lett.} \textbf{\bibinfo{volume}{92}},
  \bibinfo{eid}{107903} (\bibinfo{year}{2004}).

\end{thebibliography}
\end{document}